\RequirePackage{tikz}

\documentclass[sn-mathphys,Numbered]{sn-jnl}% Math and Physical Sciences Reference Style
\usetikzlibrary{arrows,automata}

\usepackage{graphicx}%
\usepackage{multirow}%
\usepackage{amsmath,amssymb,amsfonts}%
\usepackage{amsthm}%
\usepackage{mathrsfs}%
\usepackage[title]{appendix}%
\usepackage{xcolor}%
\usepackage{textcomp}%
\usepackage{manyfoot}%
\usepackage{booktabs}%
\usepackage{algorithm}%
\usepackage{algorithmicx}%
\usepackage{algpseudocode}%
\usepackage{listings}%
\usepackage{csquotes}
\theoremstyle{thmstyleone}%
\newtheorem{theorem}{Theorem}%  meant for continuous numbers
\newtheorem{lemma}[theorem]{Lemma}

\theoremstyle{thmstyletwo}%

\theoremstyle{thmstylethree}%

\begin{document}

\title[Directed Graph Hashing]{Directed Graph Hashing}


\author[1]{\fnm{Caleb} \sur{Helbling}}\email{caleb@purdue.edu}

\affil[1]{\orgdiv{Department of Computer Science}, \orgname{Purdue University}, \orgaddress{\street{610 Purdue Mall}, \city{West Lafayette}, \postcode{47907}, \state{Indiana}, \country{United States}}}


\abstract{This paper presents several algorithms for hashing directed graphs. The algorithms given are capable of hashing entire graphs as well as assigning hash values to specific nodes in a given graph. The notion of node symmetry is made precise via computation of vertex orbits and the graph automorphism group, and nodes that are symmetrically identical are assigned equal hashes. We also present a novel Merkle-style hashing algorithm that seeks to fulfill the recursive principle that a hash of a node should depend only on the hash of its neighbors. This algorithm works even in the presence of cycles, which would not be possible with a naive approach. Structurally hashing trees has seen widespread use in blockchain, source code version control, and web applications. Despite the popularity of tree hashing, directed graph hashing remains unstudied in the literature. Our algorithms open new possibilities to hashing both directed graphs and more complex data structures that can be reduced to directed graphs such as hypergraphs.}




\maketitle

\section{Introduction}

Hashing functions are the workhorses of the modern cryptographic stack. They are used in data storage and retrieval, as well as in digital fingerprint and checksum applications. Traditionally, hash functions take as input bit strings of arbitrary length and return a bit string of a fixed size $\mathcal{H} : \{0,1\}^* \rightarrow \{0,1\}^n$. Good hashing functions should be fast to compute, minimize the chance of duplication of output values, and produce large changes in output for small changes in input. Over the years, many bit string hashing functions have been created, including MD5 \cite{md5}, SHA-2 \cite{sha2}, and SHA-3 \cite{sha3}.

Hashing directed graphs is a topic of interest in computer science and mathematics because it allows for efficient representation of graph data. Many types of programmatic data structures and real world data can be represented as directed graphs. Being able to assign hash values to an entire graph as well as individual nodes in a graph would therefore be very valuable. There are a wide range of potential applications for these hash values, including database systems, molecular chemistry, and programming languages.

Hashing functions can be extended to trees through a simple recursive hashing scheme known as Merkle tree hashing \cite{merkle1987digital}. In Merkle tree hashing, the hash of a node is computed by hashing the concatenation of the node's label with the recursive hash of the node's children. Algorithm \ref{alg:merklehashing} gives a simple recursive implementation for Merkle hashing over binary trees. Merkle hashing has seen use in many applications, notably cryptocurrency \cite{bitcoin} and version control systems such as git \cite{git}.

\begin{algorithm}
\caption{Recursive Merkle-tree Hashing of Binary Tree}
\[
\text{mh}(n) =
\begin{cases} 
    \mathcal{H}(n.\text{label}) & n \text{ is a leaf} \\
    \mathcal{H}(n.\text{label} \text{\textbar\textbar} \text{mh}(n.\text{left}) \text{\textbar\textbar} \text{mh}(n.\text{right})) & \text{otherwise}
\end{cases}
\]
\label{alg:merklehashing}
\end{algorithm}

Merkle tree hashing can also be used to hash directed acyclic graphs (DAGs). The hashing works the same as in the tree case, and any cases of diamond shaped dependencies (as in Figure \ref{fig:diamondproblem}) are implicitly expanded to an equivalent tree. Many applications can tolerate this implicit tree expansion, which makes Merkle DAG hashing useful in practice.

\begin{figure}  
\centering 
\begin{tikzpicture}[
    > = stealth, % arrow head style
    shorten > = 0pt, % don't touch arrow head to node
    auto,
    node distance = 1.5cm, % distance between nodes
    semithick % line style
]
\tikzstyle{every state}=[
    draw = black,
    thin,
    fill = white,
    minimum size = 6mm
]

\node[state] (a) {a};
\node[state] (b) [below left of=a] {b};
\node[state] (c) [below right of=a] {c};
\node[state] (d) [below right of=b] {d};

\path[->] (a) edge node {} (b);
\path[->] (a) edge node {} (c);
\path[->] (b) edge node {} (d);
\path[->] (c) edge node {} (d);
\end{tikzpicture}
\begin{tikzpicture}[
    > = stealth, % arrow head style
    shorten > = 0pt, % don't touch arrow head to node
    auto,
    node distance = 1.5cm, % distance between nodes
    semithick % line style
]
\tikzstyle{every state}=[
    draw = black,
    thin,
    fill = white,
    minimum size = 6mm
]

\node[state] (a) {a};
\node[state] (b) [below left of=a] {b};
\node[state] (c) [below right of=a] {c};
\node[state] (d1) [below of=b] {d};
\node[state] (d2) [below of=c] {d};

\path[->] (a) edge node {} (b);
\path[->] (a) edge node {} (c);
\path[->] (b) edge node {} (d1);
\path[->] (c) edge node {} (d2);
\end{tikzpicture}
\caption{The Merkle DAG hashes of the two nodes labelled `a' are identical for these two graphs despite the fact that the graphs are not isomorphic.}
\label{fig:diamondproblem}
\end{figure}

A Merkle-style hashing technique can not be naively extended to directed graphs with cycles. Any cycle would lead to infinite recursion, due to the requirement that the hash of a node depends on the hash of its children. In this paper we work around this issue by operating on the condensation graph. The condensation graph of some graph $G$ is formed by contracting all nodes in a strongly connected component into a single node. Strongly connected components are hashed as a whole by utilizing a graph canonization algorithm. The condensation graph is a DAG by construction, which makes it amenable to Merkle DAG hashing. Combining these two concepts leads to an Merkle-style algorithm for directed graphs with cycles.

\section{Background}

Understanding the graph hashing algorithms presented in this paper requires some background in basic graph theory. The algorithm heavily utilizes analysis of the strongly connected components of a directed graph as well as the induced partition subgraph, known as the condensation graph. A \textit{strongly connected component} of a graph is a maximal set of vertices such that there exists a path in the graph between any two vertices within the strongly connected component. A graph can contain many strongly connected components (SCCs).

\begin{figure}
  \centering
  \includegraphics[width=0.5\linewidth]{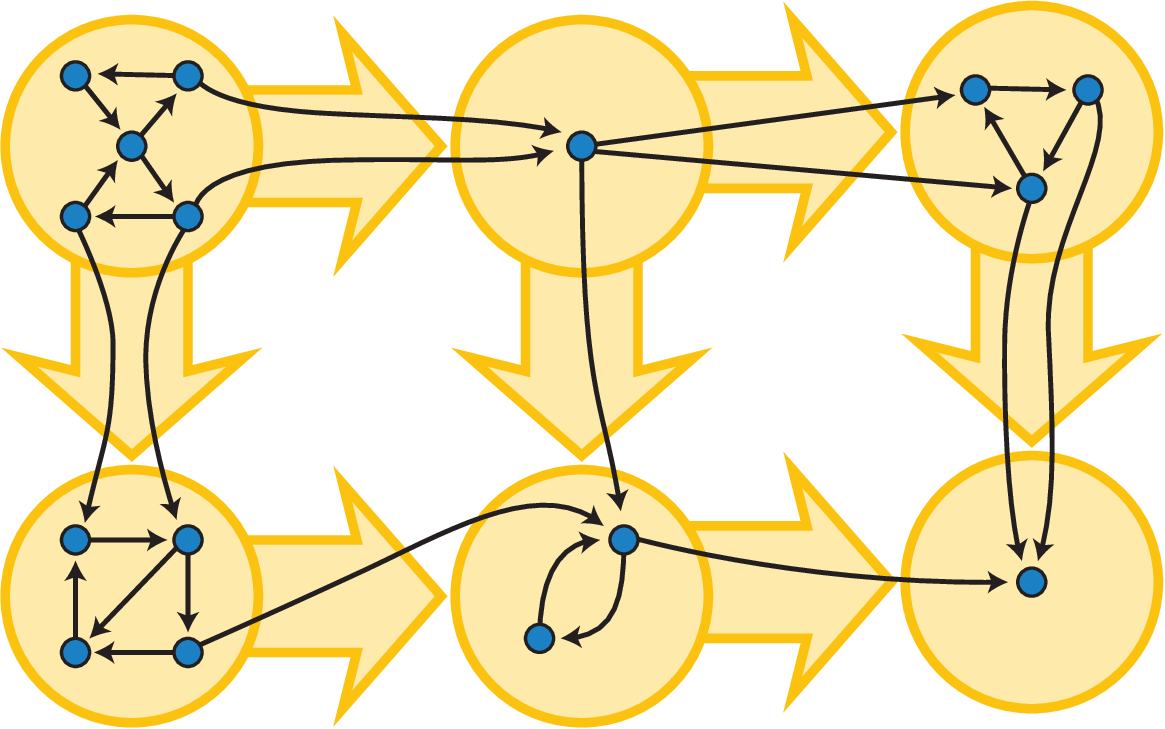}
  \caption{Each yellow node contains a strongly connected component of a graph. When each strongly connected component is contracted into a single node, the yellow condensation graph is formed.}
  \label{fig:condensation}
\end{figure}

A \textit{condensation graph} is formed by contracting strongly connected components into single nodes, and it is acyclic by construction. Figure \ref{fig:condensation} shows a graphical illustration of strongly connected components as well as the condensation graph.

An \textit{automorphism} of some graph $G=(V,E,L)$ is defined as a permutation $\sigma$ of vertices such that $(u, v) \in E$ if and only if $(\sigma(u), \sigma(v)) \in E$. Informally, the automorphism group gives all of the ways that $G$ is symmetrical.

\[
\text{Aut}(G) = \{ \sigma : V \rightarrow V \mid (u, v) \in E \leftrightarrow (\sigma(u), \sigma(v)) \in E\}
\]

A \textit{vertex orbit} is the equivalence class of vertices of a graph under the action of automorphisms. Informally, the orbit of a vertex $v$ can be thought of as the set of vertices that are symmetrically identical to $v$.

\[
\text{Orb}(v,G) = \{u \in V \mid \exists \sigma \in \text{Aut}(G) \text{ such that } \sigma(v) = u\}
\]

\textit{Canonical labeling} is the process of placing ordered vertex labels on a graph such that graphs that are isomorphic become identical after the canonization process. A canonical label for a graph is unique for a graph up to automorphism. A canonical labeling algorithm cannot distinguish between different automorphisms, because the graph under an automorphism is isomorphic to the original graph. Canonical labeling algorithms are widely used in practice for graph isomorphism testing. Examples of canonical labelling algorithms include nauty \cite{nauty}, bliss \cite{bliss}, and saucy \cite{saucy}.

\section{Algorithm}

\subsection{Graph Hashing}

\begin{figure}[t]  
\centering 
\begin{tikzpicture}[
    > = stealth, % arrow head style
    shorten > = 0pt, % don't touch arrow head to node
    auto,
    node distance = 1.5cm, % distance between nodes
    semithick % line style
]
\tikzstyle{every state}=[
    draw = black,
    thin,
    fill = white,
    minimum size = 6mm
]

\node[state] (a) {};
\node[state] (b) [below left of=a] {};
\node[state] (c) [below right of=a] {};
\node[state] (f) [right of=c] {};
\node[state] (d) [above of=f] {};
\node[state] (e) [right of=d] {};
\node[state] (g) [below of=e] {};

\path[->] (a) edge node {} (b);
\path[->] (b) edge node {} (c);
\path[->] (c) edge node {} (a);

\path[->] (d) edge node {} (e);
\path[->] (e) edge node {} (g);
\path[->] (g) edge node {} (f);
\path[->] (f) edge node {} (d);
\end{tikzpicture}
\caption{The nodes in this triangular graph should hash to the same value because they are all in the same orbit. The nodes in the square graph should hash to the same value for the same reason.}
\label{fig:trianglegraph}
\end{figure}

We now distinguish between creating a hash for a graph and hash for a node in a graph. The hash of a graph should be a unique fingerprint of an entire graph, whereas the hash of a node should be a fingerprint of a specific node in a specific graph. One property that we want to preserve is equality of hashes over isomorphism. That is, $\mathcal{H_G}(G_1)=\mathcal{H_G}(G_2)$ if and only if $G_1 \cong G_2$. Extending this idea to nodes, $\mathcal{H_N}(v_1, G_1)=\mathcal{H_N}(v_2, G_2)$ if and only if $G_1 \cong G_2$, $\exists \phi . v_2 \in \text{Orb}(\phi(v_1), G_2)$, and $\exists \phi . v_1 \in \text{Orb}(\phi^{-1}(v_2), G_1)$ where $\phi : V_1 \rightarrow V_2$ is an isomorphism from $G_1$ to $G_2$.

To achieve these properties, we build off the string hashing functions and canonical labeling algorithms. Hashing an entire graph is straightforward and begins by canonizing the entire digraph. The adjacency list (sorted using the canonical order) and the vertex labels (ordered using the canonical order) are represented as strings and hashed using a string hashing function. Algorithm \ref{digraphhash} gives an implementation of this approach. Note that the return value of the canonize function is a mapping from the vertices of the input to the canonical order.

\begin{algorithm}
    \caption{Directed Graph Hash}\label{digraphhash}
    \begin{algorithmic}[1]
    \Function{$\mathcal{H_G}$}{G=(V,E,L)}
        \State \Comment{V is the set of vertices, E is the set of edges, L is the node labeling function}
        \State $\rho \gets$ \textproc{canonize}($G$)
        \State adj\_list $\gets$ \textproc{sort}($\{(\rho(u), \rho(v)) \mid (u,v) \in E\}$)
        \State labels $\gets$ [L($\rho^{-1}(n)$) \textbf{for} n \textbf{in} 1..\textbar V\textbar]
        \State \Return $\mathcal{H}$(\textproc{to\_str}((labels, adj\_list)))
    \EndFunction
    \end{algorithmic}
\end{algorithm}

\begin{theorem}
\label{thm:digraphhashcorrect}
$\mathcal{H_G}(G_1)=\mathcal{H_G}(G_2)$ if and only if $G_1 \cong G_2$.
\end{theorem}

\begin{proof}
($\Rightarrow$)If we consider the string hashing function to be injective, then if $\mathcal{H_G}(G_1)=\mathcal{H_G}(G_2)$, then the canonized adjacency list/vertex label lists are identical. If these representations of $G_1$ and $G_2$ are identical, then $G_1$ and $G_2$ have identical canonical form $G_3$. This means that $G_1 \cong G_3$ and $G_2 \cong G_3$, which implies that $G_1 \cong G_2$ by transitivity. Of course real string hashing functions are not injective by the pigeonhole principle, however in practice nobody has yet found collisions for algorithms such as SHA512. Therefore we treat string hashing algorithms as injective, even though this isn't strictly true. The identity function or an asymmetric encryption algorithm are examples of truly injective string transformation functions, so those could be used as alternatives if desired.

($\Leftarrow)$ Let $G_3$ be the output graph of canonizing $G_1$, and $G_4$ be the output of canonizing graph $G_2$. Since $G_1 \cong G_2$, we conclude that $G_3=G_4$ since isomorphic graphs canonize to the same graph. This implies that the sorted adjacency lists and ordered vertex label lists are equivalent, so the string representation is equivalent. Therefore the string hash output are equivalent as well, which shows that $\mathcal{H_G}(G_1)=\mathcal{H_G}(G_2)$.
\end{proof}

The algorithm for hashing nodes in a graph builds on the algorithm for hashing graphs. First, the graph is canonized and the hash of the entire canonical graph is computed using Algorithm \ref{digraphhash}. Then we compute the orbit index of the input vertex, which is defined to be the canonical index of the equivalence class representative element, where the equivalence relation in this case is the vertex orbit. The orbit index in a sense creates a ``pointer'' to a specific orbit of the input graph. Figure \ref{fig:trianglegraph} demonstrates an example of this concept, and illustrates how nodes in the same orbit should be hashed to the same value. The orbit index and the graph hash are then hashed using the string hashing function.

\begin{algorithm}
    \caption{Directed Graph Node Hash}\label{digraphnodehash}
    \begin{algorithmic}[1]
    \Function{$\mathcal{H_N}$}{v, G}
        \State $\rho \gets$ \textproc{canonize}(G)
        \State g\_hash $\gets \mathcal{H_G}(G)$
        \State orbit\_idx $\gets$ $\min(\text{Orb}(\rho(v),\rho(G)))$
        \State \Return $\mathcal{H}$(\textproc{to\_str}((orbit\_idx, g\_hash)))
    \EndFunction
    \end{algorithmic}
\end{algorithm}

\begin{theorem}
\label{thm:digraphnodehashcorrect}
$\mathcal{H_N}(v_1, G_1)=\mathcal{H_N}(v_2, G_2)$ if and only if $G_1 \cong G_2$,\\
$\exists \phi . v_2 \in \text{Orb}(\phi(v_1), G_2)$, and $\exists \phi . v_1 \in \text{Orb}(\phi^{-1}(v_2), G_1)$
\end{theorem}

\begin{proof}
($\Rightarrow$) If $\mathcal{H_N}(v_1, G_1)=\mathcal{H_N}(v_2, G_2)$, then the orbit index of $v_1$ in $\mathcal{H_N}(v_1, G_1)$ is equal to the orbit index of $v_2$ in $\mathcal{H_N}(v_2, G_2)$, and $\mathcal{H_G}(G_1)=\mathcal{H_G}(G_2)$ by the injectivity of string hashing. Then by Theorem \ref{thm:digraphhashcorrect}, $G_1 \cong G_2$.

Let $\rho_1$ be the canonization of $G_1$ and $\rho_2$ be the canonization of $G_2$. Then $\rho_1(G_1)=\rho_2(G_2)$ since $G_1 \cong G_2$. Note that $\rho_1$ and $\rho_2$ are both isomorphisms. Then let $\phi=\rho_2^{-1} \circ \rho_1$ and $\phi^{-1}=\rho_1^{-1} \circ \rho_2$. Applying this definition of $\phi$, we wish to show that $v_2 \in \text{Orb}(\rho_2^{-1}(\rho_1 (v_1)), G_2)$ and $v_1 \in \text{Orb}(\rho_1^{-1}(\rho_2 (v_2)), G_1)$. Applying the isomorphisms $\rho_2$ and $\rho_1$ to these sets respectively, this is equivalent to showing that $\rho_2(v_2) \in \text{Orb}(\rho_1 (v_1), \rho_2(G_2))$ and $\rho_1(v_1) \in \text{Orb}(\rho_2 (v_2), \rho_1(G_1))$.

Recall that the orbit index of $v_1$ in $\mathcal{H_N}(v_1, G_1)$ is equal to the orbit index of $v_2$ in $\mathcal{H_N}(v_2, G_2)$. This implies that the two orbit sets are equal $\text{Orb}(\rho_1 (v_1), \rho_1(G_1))=\text{Orb}(\rho_2 (v_2), \rho_2(G_2))$ since they have equal representative elements. Then since $\rho_1(G_1)=\rho_2(G_2)$, $\text{Orb}(\rho_1 (v_1), \rho_1(G_1))=\text{Orb}(\rho_2 (v_2), \rho_2(G_2))=\text{Orb}(\rho_1 (v_1), \rho_2(G_2))$\\
$=\text{Orb}(\rho_2 (v_2), \rho_1(G_1))$. Then trivially $\rho_1(v_1) \in \text{Orb}(\rho_1 (v_1), \rho_1(G_1))$ and $\rho_2(v_2) \in \text{Orb}(\rho_2 (v_2), \rho_2(G_2))$. Following the chain of equalities and making the appropriate substitutions, then $\rho_2(v_2) \in \text{Orb}(\rho_1 (v_1), \rho_2(G_2))$ and $\rho_1(v_1) \in \text{Orb}(\rho_2 (v_2), \rho_1(G_1))$.

($\Leftarrow$) If $G_1 \cong G_2$, then $\mathcal{H_G}(G_1)=\mathcal{H_G}(G_2)$ by Theorem \ref{thm:digraphhashcorrect}. Then there exists some isomorphism $\phi$ and $\phi'$ such that $v_2 \in \text{Orb}(\phi(v_1),G_2)$ and $v_1 \in \text{Orb}(\phi'^{-1}(v_2),G_1)$. Let $\rho_1$ be the canonization of $G_1$ and $\rho_2$ be the canonization of $G_2$. Applying these isomorphisms, we have $\rho_2(v_2) \in \text{Orb}(\rho_2(\phi(v_1)), \rho_2(G_2))$ and $\rho_1(v_1) \in \text{Orb}(\rho_1(\phi'^{-1}(v_2)),\rho_1(G_1))$. Let $f=\rho_1 \circ \phi^{-1} \circ \rho_2^{-1}$ and $f'=\rho_2 \circ \phi' \circ \rho_1^{-1}$ be automorphisms over the canonized graph.

Then by definition of orbit, $\text{Orb}(\rho_2(\phi(v_1)), \rho_2(G_2))=\text{Orb}(f(\rho_2(\phi(v_1))), \rho_2(G_2))$ and $\text{Orb}(\rho_1(\phi'^{-1}(v_2)),\rho_1(G_1)) = \text{Orb}(f'(\rho_1(\phi'^{-1}(v_2)),\rho_1(G_1)))$. Substituting in the definitions of $f$ and $f'$, we have $\text{Orb}(\rho_2(\phi(v_1)), \rho_2(G_2))=\text{Orb}(\rho_1(v_1), \rho_2(G_2))$ and $\text{Orb}(\rho_1(\phi'^{-1}(v_2)),\rho_1(G_1)) = \text{Orb}(\rho_2(v_2),\rho_1(G_1))$. This shows that $\rho_2(v_2) \in \text{Orb}(\rho_1(v_1), \rho_2(G_2))$ and $\rho_1(v_1) \in \text{Orb}(\rho_2(v_2),\rho_1(G_1))$. Since $G_1 \cong G_2$, $\rho_1(G_1)=\rho_2(G_2)$ so $\rho_2(v_2) \in \text{Orb}(\rho_1(v_1), \rho_1(G_1))$ and $\rho_1(v_1) \in \text{Orb}(\rho_2(v_2),\rho_2(G_2))$. This suffices to show that the orbit indices of are equal in the computations of $\mathcal{H_N}(v_1, G_1)$ and $\mathcal{H_N}(v_2, G_2)$.

Since the orbit indices are equal and $\mathcal{H_G}(G_1)=\mathcal{H_G}(G_2)$, then $\mathcal{H_N}(v_1, G_1)=\mathcal{H_N}(v_2, G_2)$.
\end{proof}

\subsection{Quotient Graph Hashing}

\begin{figure}  
\centering 
\begin{tikzpicture}[
    > = stealth, % arrow head style
    shorten > = 0pt, % don't touch arrow head to node
    auto,
    node distance = 1.5cm, % distance between nodes
    semithick % line style
]
\tikzstyle{every state}=[
    draw = black,
    thin,
    fill = white,
    minimum size = 6mm
]

\node[state] (a) {};

\path (a) edge [loop above] node {} (a);
\end{tikzpicture}
\caption{The quotient graph G/Orb of the graphs in Figure \ref{fig:trianglegraph}. Both graphs have the same quotient graph.}
\label{fig:quotientgraph}
\end{figure}

The algorithms in the previous section utilize the hash of an entire graph $G$ via hashing the adjacency list of $G$. To explore the design space of possible graph hashing algorithms, consider the case where we hash the quotient graph $G/Orb$ instead of $G$. The quotient graph is defined as the graph constructed from contracting nodes in the same orbit into a single node, while preserving the inter-orbit connections. More formally, $G/Orb=(V',E')$ is a multi-edge graph (ie, $E'$ is a multiset) defined as follows:

\begin{equation}
V'=\{\text{Representative}(\text{Orb}(x,G)) \mid x \in V\}
\end{equation}

\begin{equation}
E'=[(u,\text{Representative}(\text{Orb}(v,G))) \mid u \in V' \land (u,v) \in E]
\end{equation}

The Representative function simply selects a single representative node from each orbit. Since all nodes in the same orbit are symmetrically identical, choosing any node from each orbit will suffice. Hashing a multi-edge graph can be done by performing a reduction to a directed graph and then using Algorithm \ref{digraphhash}. Section \ref{sec:extensions} gives a brief overview of how this encoding can be done with only a small overhead.

\begin{figure}
    \centering
    \includegraphics[scale=0.75]{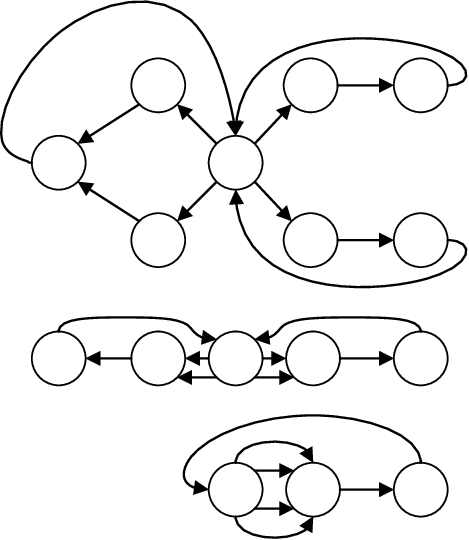}
    \caption{Let $G$ be the top graph. Then the middle graph is a depiction of $G/Orb$, and the bottom graph is a depiction of $(G/Orb)/Orb$.}
    \label{fig:multiquotientgraph}
\end{figure}

Extending the node hashing algorithm is not as straightforward. Algorithm \ref{digraphnodehash} creates an orbit index into the original graph $G$ and not $G/Orb$. An obvious workaround to computing an index into $G/Orb$ is to simply canonize $G/Orb$ and use the canonical label as the index. This will only work if $|Aut(G/Orb)|=1$, that is $G/Orb$ eliminates all symmetry from $G$ and the canonical labeling algorithm only has one choice of index label for each node. Unfortunately this is not the case, as the counterexample in Figure \ref{fig:multiquotientgraph} demonstrates. To workaround this issue, we can simply take the quotient of the graph recursively $((G/Orb)/Orb)/Orb ... $ until a fixed point is reached. Algorithm \ref{alg:quotientfixpoint} gives an implementation of this procedure. Note that $\sigma$ refers to a function that maps nodes in $G$ to nodes in $G/Orb$.

\begin{algorithm}
    \caption{Directed Graph Node Hash}\label{alg:quotientfixpoint}
    \begin{algorithmic}[1]
    \Function{QuotientFixpoint}{$G$}
        \State $(\sigma,G') \gets G/Orb$ \Comment{$\sigma$ maps nodes in $G$ to nodes in $G'$}
        \If {$|G'|=|G|$}
            \State \Return $(\sigma,G')$
        \Else
            \State $(\sigma',G'')\gets$ \textproc{QuotientFixpoint}($G'$)
            \State \Return $(\sigma' \circ \sigma,G'')$
        \EndIf
    \EndFunction
    \end{algorithmic}
\end{algorithm}

\begin{lemma}
\label{thm:qfterminates}
For all input graphs $G$, \textproc{QuotientFixpoint($G$)} terminates.
\end{lemma}

\begin{proof}
Proceed by performing strong induction on the number of nodes in $G$. As a base case, consider the empty graph $|G|=0$. Then $G/Orb$ is also the empty graph, so $|G|=|G/Orb|$. \textproc{QuotientFixpoint} immediately returns in this case.

For the inductive step, assume \textproc{QuotientFixpoint} terminates on graphs of size $0,1,...,n-1$. Then we wish to show that \textproc{QuotientFixpoint} terminates on a graph of size $|G|=n$. Since $G/Orb$ is a quotient graph, $|G/Orb| \leq |G|$. In the case where $|G/Orb|=|G|$, \textproc{QuotientFixpoint} immediately terminates. In the case where $|G/Orb| < |G|$, then the recursive call terminates by the inductive hypothesis. This implies that \textproc{QuotientFixpoint} terminates on a graph of size $n$.
\end{proof}

One point of interest is whether or not all symmetry from the graph is fully eliminated by this procedure:\\

\begin{lemma}
\label{thm:qfsize1}
For all graphs $G$, $|Aut(G')|=1$ where $(\sigma,G')=\text{\textproc{QuotientFixpoint}}(G)$
\end{lemma}

\begin{proof}
There are two exit points in the \textproc{QuotientFixpoint} function. In the second return case, the return value is equal to what a recursive call to \textproc{QuotientFixpoint} returns. In the first return case, we have the constraint $|G'/Orb|=|G'|$. This implies that the size of all orbits in $G'$ and $|G'/Orb|$ are 1. By the definition of orbit, this means that $|Aut(G')|=1$ (that is the identity function is the only function in $Aut(G')$). Since \textproc{QuotientFixpoint} terminates by Lemma \ref{thm:qfterminates}, the return value of the overall function is determined by the first return case. Therefore $|Aut(\text{\textproc{QuotientFixpoint}}(G))|=1$.
\end{proof}

\begin{algorithm}
    \caption{Directed Quotient Graph Hash}\label{alg:digraphhashqf}
    \begin{algorithmic}[1]
    \Function{$\mathcal{H_{GQF}}$}{G}
        \State $(\sigma,G') \gets$ \textproc{QuotientFixpoint}(G)
        \State $G'' \gets$ \textproc{ReduceToDigraph}($G'$) \Comment{$G'$ is a multi-edge digraph, so we need to convert it to a digraph}
        \State \Return $\mathcal{H_G}(G'')$
    \EndFunction
    \end{algorithmic}
\end{algorithm}

Algorithm 4 gives the procedure for hashing the graph after the quotient fixpoint has been reached. After computing the quotient fixpoint, the multi-edged graph is reduced to an isomorphic directed graph and Algorithm \ref{digraphhash} is applied.\\

\begin{theorem}
$\mathcal{H_{GQF}}(G_1)=\mathcal{H_{GQF}}(G_2)$ if and only if $G_1' \cong G_2'$ where $(\sigma_1,G_1')=\textproc{QuotientFixpoint}(G_1)$ and $(\sigma_2,G_2')=\textproc{QuotientFixpoint}(G_2)$.
\end{theorem}

\begin{proof}
($\Rightarrow$) Let $G''$ refer to the $G''$ graph in Algorithm \ref{alg:digraphhashqf}. If $\mathcal{H_{GQF}}(G_1)=\mathcal{H_{GQF}}(G_2)$, then $G''_1 \cong G''_2$ by Theorem \ref{thm:digraphhashcorrect}. Since \textsc{ReduceToDigraph} is an isomorphic reduction, then $G_1' \cong G_2'$.

($\Leftarrow$) Suppose that $G'_1 \cong G'_2$. Since \textsc{ReduceToDigraph} is an isomorphic reduction, then $G''_1 \cong G''_2$. By Theorem \ref{thm:digraphhashcorrect}, $\mathcal{H_G}(G''_1)=\mathcal{H_G}(G''_2)$. This directly shows that $\mathcal{H_{GQF}}(G_1)=\mathcal{H_{GQF}}(G_2)$.
\end{proof}

Algorithm \ref{alg:digraphhashnqf} gives a procedure for hashing a node in the quotient fixpoint graph. Similar to Algorithm \ref{alg:digraphhashqf}, the quotient fixpoint is computed and reduced to a digraph, then Algorithm \ref{digraphnodehash} can be applied.\\

\begin{algorithm}
    \caption{Directed Quotient Graph Node Hash}\label{alg:digraphhashnqf}
    \begin{algorithmic}[1]
    \Function{$\mathcal{H_{NQF}}$}{v,G}
        \State $(\sigma,G') \gets$ \textproc{QuotientFixpoint}($G$)
        \State $G'' \gets$ \textproc{ReduceToDigraph}($G'$)
        \State \Return $\mathcal{H_N}(\sigma(v), G'')$
    \EndFunction
    \end{algorithmic}
\end{algorithm}

\begin{theorem}
\label{thm:hnqfcorrect}
$\mathcal{H_{NQF}}(v_1, G_1)=\mathcal{H_{NQF}}(v_2, G_2)$ if and only if $G_1'$ $\cong$ $G_2'$, $\exists \phi . \sigma_2(v_2) = \phi(\sigma_1(v_1))$, and $\exists \phi . \sigma_1(v_1) = \phi^{-1}(\sigma_2(v_2))$ where $(\sigma_1,G_1')=\text{\textproc{QuotientFixpoint}}(G_1)$,\\
$(\sigma_2,G_2')=\text{\textproc{QuotientFixpoint}}(G_2)$, and $\phi$ is an isomorphism from $G_1'$ to $G_2'$.
\end{theorem}

\begin{proof}
Since \textproc{ReduceToDigraph} is an isomorphism, this proposition directly follows from Theorem \ref{thm:digraphnodehashcorrect}, Lemma \ref{thm:qfsize1}, and the implementation of Algorithm \ref{alg:digraphhashnqf} by making the appropriate substitutions and keeping in mind that all orbits in $G_1'$ and $G_2'$ have size 1.
\end{proof}

\subsection{Merkle Graph Hashing}

The Merkle hash property that a hash of a node should only depends on the hash of its neighbors is desirable for some applications, particularly applications where graphs are built up over time. The algorithms presented so far all require canonization passes over the entire graph, which in some situations is too expensive. Therefore we now investigate how we can extend Merkle hashing to directed graphs.

By combining the condensation graph, canonical labeling procedure and the Merkle hashing algorithm for trees, we arrive at an algorithm that can recursively hash directed graphs. Algorithm \ref{alg:merklegraphhash} gives the pseudocode for this approach. The algorithm begins by creating a condensation graph which is acyclic by construction. Then each strongly connected component (a node in the condensation graph) is hashed. When hashing a SCC, the successor SCCs in the condensation graph are recursively hashed first. The algorithm then proceeds to compute new labels for each node in the SCC by concatenating to the labels a sorted list of the hashes of the non-SCC successors. Then the final hash for the node is computed using Algorithm \ref{digraphnodehash} over the relabeled SCC subgraph.\\

\begin{algorithm}
    \caption{Merkle Graph Node Hashing}\label{alg:merklegraphhash}
    \begin{algorithmic}[1]
    \Function{$\mathcal{H_{NM}}$}{v,G=(V,E,L)}
        \State $S \gets$ \textproc{Scc}(v,$G$) \Comment{$S$ is the set of vertices in the SCC containing v}
        \State $L'(u)=\textproc{to\_str}((L(u),\textproc{sort}([\mathcal{H_{NM}}(w,G) \mid u \in S \land w \notin S \land (u,w) \in E])))$
        \State $(V', E', L'') \gets$ \textproc{Subgraph}(S, $G$)
        \State $G' \gets (V',E',L')$
        \State \Return $\mathcal{H_N}(v, G')$ \Comment{Hash the relabelled SCC containing v}
    \EndFunction
    \end{algorithmic}
\end{algorithm}

\begin{theorem}
For all inputs $v,G$, $\mathcal{H_{NM}}(v,G)$ terminates.
\end{theorem}

\begin{proof}
Proceed by performing strong induction over the number of strongly connected components in $G$ reachable from $v$. As a base case, consider $n=1$ strongly connected components are reachable from $v$. Then in the relabeling procedure $L'$, there are no nodes $w$ satisfying the constraint $w \notin Scc(v,G)$ since the only SCC reachable from a node in $Scc(v,G)$ is another node in $Scc(v,G)$. Therefore $\mathcal{H_{NM}}$ is never recursively called.

For the inductive step, assume $\mathcal{H_{NM}}$ terminates on cases $1...n-1$ strongly connected components reachable from the input node. Let $n$ be the number of strongly connected components reachable from $v$. Then when $L'$ is called, any $w$ which satisfies the constraint $w \notin \textproc{Scc}(v,G)$ must come from deeper in the condensation graph. This implies that the number of strongly connected components reachable from $w$ is strictly less than $n$. By the inductive hypothesis, these recursive calls terminate. Therefore $\mathcal{H_{NM}}$ terminates on all inputs.
\end{proof}

\begin{figure}
    \centering
    \includegraphics[width=0.8\textwidth]{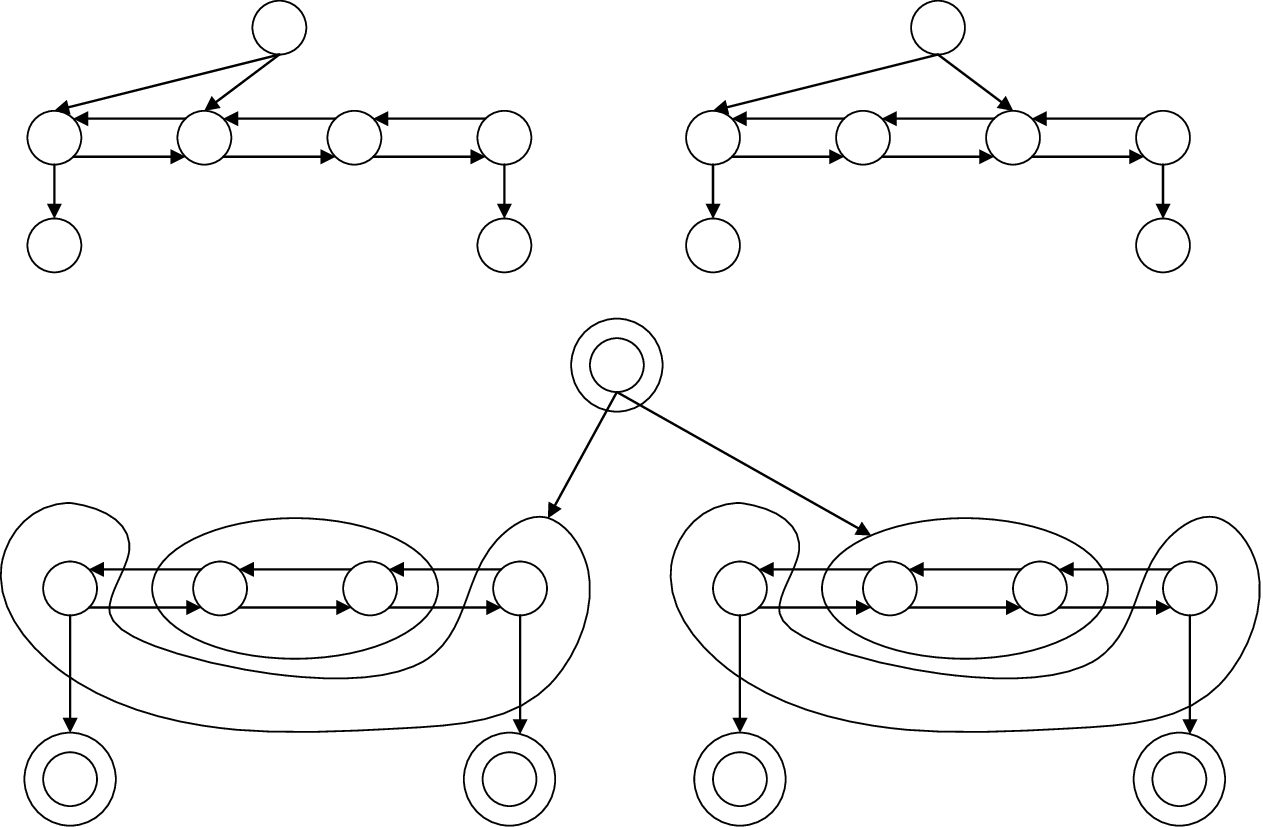}
    \caption{Two non-isomorphic digraphs (above) both expand to the same generalized hypergraph (below) as a result of the \textproc{ToMerkleDigraph} procedure. In the Merkle digraph hashing procedure, nodes in the same hyperedge hash to the same value.}
    \label{fig:hypergraphs}
\end{figure}

To make a correctness proof for this algorithm requires an additional conversion step, analogous to how \textproc{QuotientFixpoint} is used in Theorem \ref{thm:hnqfcorrect}. The conversion step \textproc{ToMerkleDigraph} takes the directed graph as input, determines the condensation of the graph, expands the condensation directed acyclic graph into a tree, then beginning from the leaves recursively creates hyperedges for each orbit in a SCC, and redirects cross-SCC edges to point to the new orbit hyperedges. The result of this conversion is a generalized hypergraph. Figure \ref{fig:hypergraphs} illustrates this conversion process.\\

\begin{algorithm}
    \caption{Convert Digraph to Merkle Digraph. This pseudocode gives an outline of the conversion process. It is possible to write a recursive formulation of this procedure.}\label{alg:merkledigraph}
    \begin{algorithmic}[1]
    \Function{ToMerkleDigraph}{G}
        \State $G' \gets$ \textproc{CondensationToTree}($G$) \Comment{Expand condensation DAG to a tree. Structures within a SCC are preserved.}
        \State $G'' \gets$ \textproc{RedirectCrossSccEdges}($G'$) \Comment{For edges crossing from one SCC to another, redirect the target of the edge to point to the orbit hyperedge instead}
        \State \Return $G''$
    \EndFunction
    \end{algorithmic}
\end{algorithm}

\begin{lemma}
\label{thm:hnmcorrectlemma}
On input $v,G$, the $G'$ computed in $\mathcal{H_{NM}}$ has an isomorphism to  $\text{\textproc{ToMerkleDigraph}}(\textproc{Subgraph}(\textproc{Reachable}(v, G), G))$
\end{lemma}

\begin{proof}
Proceed by performing strong induction over the number of strongly connected components in $G$ reachable from $v$. As a base case, consider $n=1$ strongly connected components are reachable from $v$. Then in the relabeling procedure $L'$, there are no nodes $w$ satisfying the constraint $w \notin \textproc{Scc}(v,G)$ since the only SCC reachable from a node in $\textproc{Scc}(v,G)$ is another node in $\textproc{Scc}(v,G)$. Therefore the second element of the tuple returned by the labeling function $L'$ will be the empty list. This implies that the labeling function $L'$ is equal to $L$. Then the \textproc{Subgraph} step will simply choose the SCC of $v$, leaving $V',E'$ the same as $V,E$. Likewise, the $\text{\textproc{ToMerkleDigraph}}(\textproc{Subgraph}(\textproc{Reachable}(v, G), G))$ operation will leave its input graph unchanged since only one SCC is reachable from $v$. This suffices to show that the $G'$ computed in $\mathcal{H_{NM}}$ is isomorphic to $\text{\textproc{ToMerkleDigraph}}(\textproc{Subgraph}(\textproc{Reachable}(v, G), G))$ for $n=1$.

For the inductive step, assume the $G'$ computed in $\mathcal{H_{NM}}$ has an isomorphism to $\text{\textproc{ToMerkleDigraph}}(\textproc{Subgraph}(\textproc{Reachable}(v, G),$ $ G))$ for cases $1...n-1$ strongly connected components reachable from $v$. Let $n$ be the number of strongly connected components reachable from $v$. We observe that when $L'$ is called, any $w$ which satisfies the constraint $w \notin \textproc{Scc}(v,G)$ must come from deeper in the condensation graph. This implies that the number of strongly connected components reachable from $w$ is strictly less than $n$. Applying the inductive hypotheses, all $G'$ in recursive calls to $\mathcal{H_{NM}}$ must have an isomorphism to $\text{\textproc{ToMerkleDigraph}}(\textproc{Subgraph}(\textproc{Reachable}(w, G), G))$. When used in the context of $L'$, returning $\mathcal{H_N}(w, G')$ is therefore equivalent to redirecting cross SCC edges to point to an orbit hyperedge. Furthermore $L'$ implicitly expands any DAG into a tree due to its use of strings and encodes all cross SCC edges in the label. This suffices to show that the $G'$ computed in $\mathcal{H_{NM}}$ has an isomorphism to $\text{\textproc{ToMerkleDigraph}}(\textproc{Subgraph}(\textproc{Reachable}(v, G), G))$.
\end{proof}

\begin{theorem}
\label{thm:hnmcorrect}
$\mathcal{H_{NM}}(v_1, G_1)=\mathcal{H_{NM}}(v_2, G_2)$ if and only if\\
$G_1'$ $\cong$ $G_2'$, $\exists \phi . v_2 \in \text{Orb}(\phi(v_1), G_2')$, and $\exists \phi . v_1 \in \text{Orb}(\phi^{-1}(v_2), G_1')$ where\\
$G_1'=\text{\textproc{ToMerkleDigraph}}(\textproc{Subgraph}(\textproc{Reachable}(v_1, G_1), G_1))$,\\
$G_2'=\text{\textproc{ToMerkleDigraph}}(\textproc{Subgraph}(\textproc{Reachable}(v_2, G_2), G_2))$, and $\phi$ is an isomorphism from $G_1'$ to $G_2'$.
\end{theorem}

\begin{proof}
This follows directly from applying Lemma \ref{thm:hnmcorrectlemma} and Theorem \ref{thm:digraphnodehashcorrect}.
\end{proof}

Algorithm \ref{alg:merklegraphhash} utilizes $\mathcal{H_N}$, but it is also possible to use $\mathcal{H_{NQF}}$ instead. This leads to a similar construction as $\textproc{ToMerkleDigraph}$, but with additional steps to compute the quotient graph utilizing the orbit equivalence relation. The proof of correctness for this approach follows similar reasoning as the previous proofs, and is left as an exercise to the reader.

\subsection{Node Set Hashing}

\begin{figure}[t]  
\centering 
\begin{tikzpicture}[
    > = stealth, % arrow head style
    shorten > = 0pt, % don't touch arrow head to node
    auto,
    node distance = 1.5cm, % distance between nodes
    semithick % line style
]
\tikzstyle{every state}=[
    draw = black,
    thin,
    fill = white,
    minimum size = 6mm
]

\node[state] (a) {};
\node[state, circle, fill=black, inner sep=0pt, minimum size=2mm] (p1) [left of=a] {};
\node[state] (b) [right of=a] {};
\node[state, circle, fill=black, inner sep=0pt, minimum size=2mm] (p2) [right of=b] {};
\node[state] (c) [below of=b] {};
\node[state] (h) [below of=a] {};

\node[state, circle, fill=black, inner sep=0pt, minimum size=2mm] (p3) [right of=p2] {};
\node[state] (d) [right of=p3] {};
\node[state] (f) [below of=d] {};
\node[state] (e) [right of=d] {};
\node[state] (g) [below of=e] {};
\node[state, circle, fill=black, inner sep=0pt, minimum size=2mm] (p4) [right of=g] {};

\path[->] (a) edge node {} (b);
\path[->] (b) edge node {} (c);
\path[->] (c) edge node {} (h);
\path[->] (h) edge node {} (a);

\path[->] (d) edge node {} (e);
\path[->] (e) edge node {} (g);
\path[->] (g) edge node {} (f);
\path[->] (f) edge node {} (d);

\draw[dashed, ->] (p1) -- (a);
\draw[dashed, ->] (p2) -- (b);
\draw[dashed, ->] (p3) -- (d);
\draw[dashed, ->] (p4) -- (g);

\end{tikzpicture}
\caption{Taking a pointer (represented as the dotted edge) may break the symmetry of a graph for other pointers. In the graph on the left, we have the constraint that the two nodes we point to must be directly adjacent in the cycle. In the graph on the right, we have the constraint that the two nodes we point to must be opposite from each other in the cycle.}
\label{fig:mutliplepointers}
\end{figure}

Previous sections have assumed we hash nodes by creating a pointer to a single node in the graph. However there are use cases where we may want to select (or point to) multiple nodes in the input graph. However the mere act of pointing to one specific node may break the graph symmetry for other selections, as Figure \ref{fig:mutliplepointers} illustrates.

\begin{algorithm}
    \caption{Node Set Hashing}\label{alg:nodesethash}
    \begin{algorithmic}[1]
    \Function{$\mathcal{H_{NS}}$}{$vs$,$G=(V,E,L)$}
        \State $L'(u)=\textbf{if}\; u \in vs\; \textbf{then} \; \textproc{to\_str}((``\text{ptr}", L(u)))\; \textbf{else}\; \textproc{to\_str}((``\text{nonptr}", L(u)))$
        \State $G' \gets (V, E, L')$
        \State $f(u) = \mathcal{H_{N}}(u, G')$
        \State \Return $f$
    \EndFunction
    \end{algorithmic}
\end{algorithm}

Algorithm \ref{alg:nodesethash} gives a solution for computing the hash of a set of pointee nodes. The algorithm functions by relabelling all the nodes in the input graph, prepending the string ``ptr" to the nodes in the pointee set, and prepending ``nonptr" otherwise. At this point, Algorithm \ref{digraphnodehash} can be applied to compute the hash of individual nodes. The return result is a function that maps nodes in the pointee set to hash values.

One obvious question to ask is whether or not these ideas could be applied to the Merkle graph hashing algorithm. Instead of redirecting inter-SCC edges to point to orbit hyperedges, perhaps it would be possible to point to the set of nodes in the target SCC. Unfortunately a closer analysis reveals that such a behaviour violates the fundamental philosophy behind the Merkle hashing principle of reachability. In our formulation of Merkle hashing for directed graphs, the hash of a node should only depend on the structure of the graph reachable from that node. In a recursive formulation, breaking symmetry within an SCC due to incoming inter-SCC edges cannot be taken into account, since the sources of those inter-SCC edges are not reachable from the target SCC. Therefore it seems unlikely that a better Merkle-style hashing algorithm for directed graphs could be devised.

\section{Extensions}
\label{sec:extensions}

The graph hashing algorithms can be extended to more general graph types by creating polynomial time reductions to node-labeled directed graphs. In Section 14 of the nauty User's Guide \cite{nautyuserguide}, a procedure is outlined for efficiently encoding edge labeled directed graphs as a node labeled directed graph. The rough idea is to encode each label as a binary string and create a node layer for each position in the set of bitstrings. A given edge will only appear in layer $i$ if bit $i$ is set to $1$ in the bitstring representation of that edge's label. This encoding adds an additional $\log n$ multiplier overhead. For more details, including a hypergraph encoding procedure, we refer the reader to the nauty User's Guide.

\section{Analysis}

The proposition of Theorem \ref{thm:digraphhashcorrect} shows that the graph hashing problem is \textbf{GI}-hard. It is currently unknown whether or not graph isomorphism is in \textbf{P}, \textbf{NP}-complete, or \textbf{NP}-intermediate \cite{graphisocomplexity}. If graph isomorphism is not in \textbf{P}, then graph hashing is also not in \textbf{P}.

In addition to providing a canonization procedure, some canonization libraries also provide orbit detection \cite{nautyuserguide}. There are currently no graph canonization algorithms that run in polynomial time \cite{miyazaki1997complexity}. Since the graph hashing algorithms presented here utilize graph canonization and orbit detection, this means that the overall run time complexity of the algorithm is currently not polynomial. This is not necessarily a deal breaker, since many real world directed graphs are sparse and/or have unique node labels, which makes canonization and orbit detection trivial (ie, you can simply canonize by lexicographically sorting node labels).

Of particular interest is the Merkle-style hashing algorithm, which has a recursive structure. Unlike the other algorithms presented in this paper, this algorithm employs a clear divide and conquer strategy. We now make an analysis of the upper bound of Algorithm \ref{alg:merklegraphhash}, and show that it runs in polynomial time for the special case of unique node labels.

Let $n$ and $e$ be the maximum number of nodes and maximum number of edges in a strongly connected component in the input graph. Let $m$ be the number of strongly connected components reachable from the input node, and $f(n)$ give the run-time of the graph canonization/orbit detection algorithm. Let $p$ be the maximum number of outgoing edges from a single node crossing between strongly connected components. Comparison based sorting can run in at best $\Omega(n \log n)$ time \cite{introtoalgorithms}, and strongly connected component detection in $O(V + E)$ time \cite{tarjanscc}. We now break down the runtime into the following parts:

\begin{itemize}
    \item Canonizing a strongly connected component takes $O(f(n))$ time.
    \item Sorting the adjacency list in the $\mathcal{H_G}$ subcall (Algorithm \ref{digraphhash}) takes $O(e \log e)$ time for each strongly connected component.
    \item Sorting the cross edge hashes in the relabelling function $L'$ takes $O(np \log p)$ time for each strongly connected component.
    \item Converting the node labels to a string takes $O(n)$ time for each strongly connected component.
    \item Converting the adjacency list to a string takes $O(e)$ time for each strongly connected component.
    \item Converting the cross SCC edge labels to a string in the $L'$ function takes $O(np)$ time.
\end{itemize}

Putting this altogether, the runtime of the Merkle-graph hashing algorithm is:

\begin{gather}
    O(m (f(n) + e \log (e) + n p \log p + n + e + n p)) = \\
    O(m f(n) + m e \log e + m n p \log p)
\end{gather}

In the specialized case where node labels are unique, the canonization function can be determined by lexiographically sorting the nodes by their labels. In this case $f(n)=O(n \log n)$, which gives a polynomial runtime:

\begin{equation}
    O(m n \log n + m e \log e + m n p \log p)
\end{equation}

Instead of sorting the non-recursive labels in Algorithm \ref{alg:merklegraphhash} or sorting the adjacency list in Algorithm \ref{digraphhash}, we could use a multiset hashing algorithm instead. Such an algorithm would knock off the logarithmic factor from the running time. We refer the reader to Clarke et al. \cite{Clarke2003IncrementalMH} for an example of a multiset hashing algorithm.

In the case where node labels are identical, we must rely on a graph canonization algorithm. The fastest known canonical labeling algorithm runs in $O(\exp (n^{1/2 + O(1)}))$ time \cite{miyazaki1997complexity}, where n is the number of nodes in the graph. Since orbit partitioning is polynomially equivalent to graph isomorphism \cite{graphisodisease}, then the run-time of Merkle-graph hashing remains exponential:

\begin{equation}
    O\left(m \exp \left(n^{1/2 + O(1)}\right)\right)
\end{equation}

\section{Experimental Results}

In this section, we analyze the runtime of the presented algorithms using a Python implementation. The tests took place on a system with an Intel i7-12800H processor with 32 GB of RAM. For the canonization and orbit detection steps, we utilized the pynauty library, which is a wrapper around McKay \& Piperno's nauty library (implemented in C). To generate a random graph with $n$ nodes, we randomly chose an edge count by sampling a uniform distribution of integers in the interval $[n-1,n^2]$, then generated a random graph with the chosen number of nodes and edges. The random graphs were generated with the NetworkX function \texttt{gnm\_random\_graph}, and all nodes were given the same label. Graphs that were not weakly connected were filtered out and regenerated.

\begin{figure}
    \centering
    \includegraphics[width=0.75\textwidth]{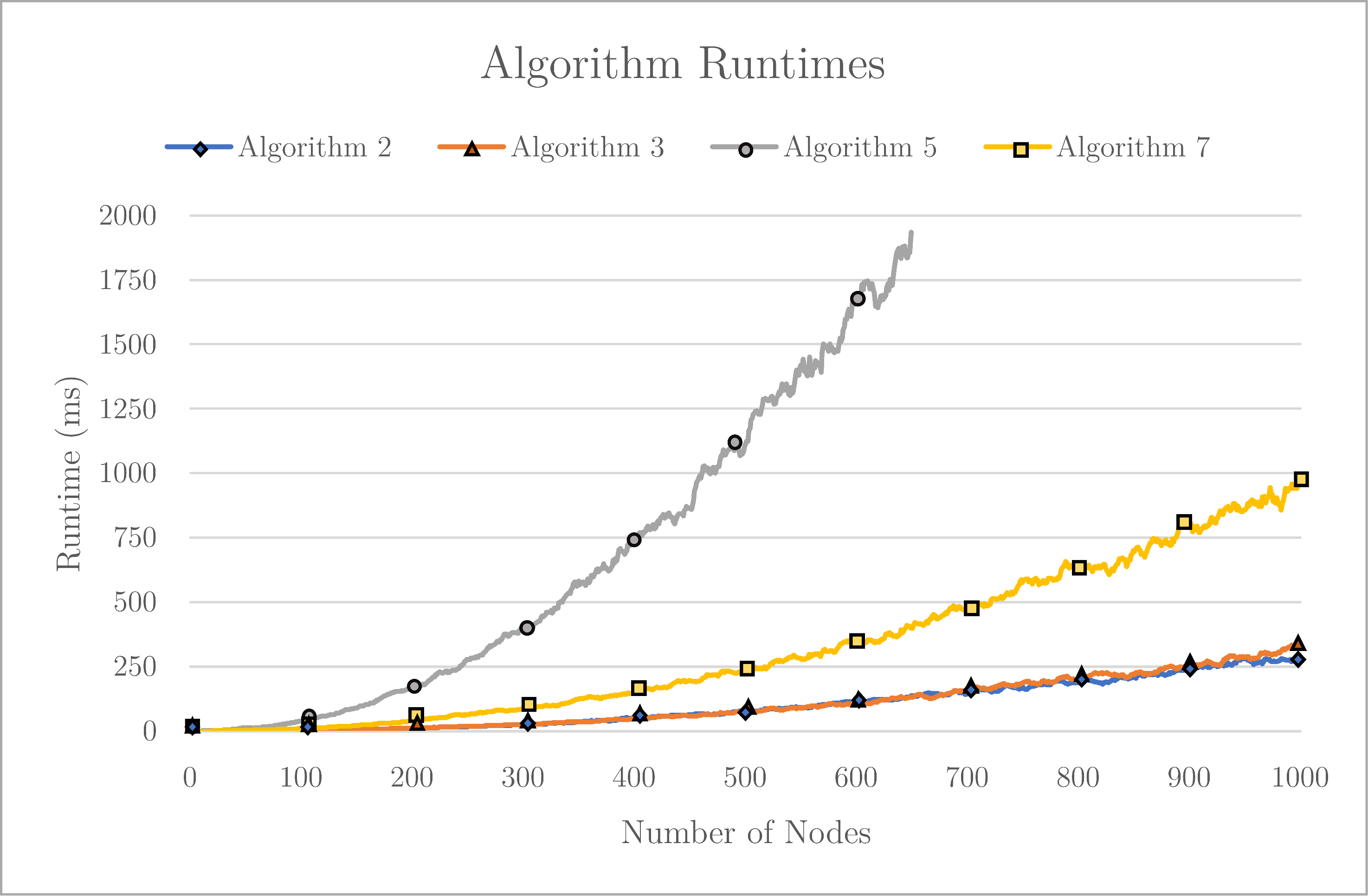}
    \caption{Median runtime of Algorithms \ref{digraphhash}, \ref{digraphnodehash}, \ref{alg:digraphhashqf} and \ref{alg:merklegraphhash} over 100 trials for $n=0$ to $1000$ nodes.}
    \label{fig:alg-all-runtime}
\end{figure}

Figure \ref{fig:alg-all-runtime} illustrates the runtimes of the algorithms presented in this paper over a range of $n=0$ to 1000 nodes. For each $n$, 100 trials were timed and the median duration was computed. As the plot illustrates, small graphs ($\leq$ 100 nodes) only take a few milliseconds to complete, and larger graphs take hundreds to thousands of milliseconds to complete. The median was used in favor of the mean due to the presence of large outliers in the runtime.

Although the general runtime of the algorithms presented in this paper are exponential, Neuen \& Schweizer \cite{Neuen2017-ft} note that it is difficult to find input graphs that lead to exponential runtime:

\begin{displayquote}
``The state-of-the-art solvers for graph isomorphism (e.g. bliss, nauty/traces, conauto, saucy, etc.) can readily solve generic instances with tens of thousands of vertices. Indeed, experiments show that on inputs without particular combinatorial structure the algorithms scale almost linearly. In fact, it is non-trivial to create challenging instances for such solvers and the number of difficult benchmark graphs available is quite limited.''
\end{displayquote}

\begin{figure}
    \centering
    \includegraphics[width=0.75\textwidth]{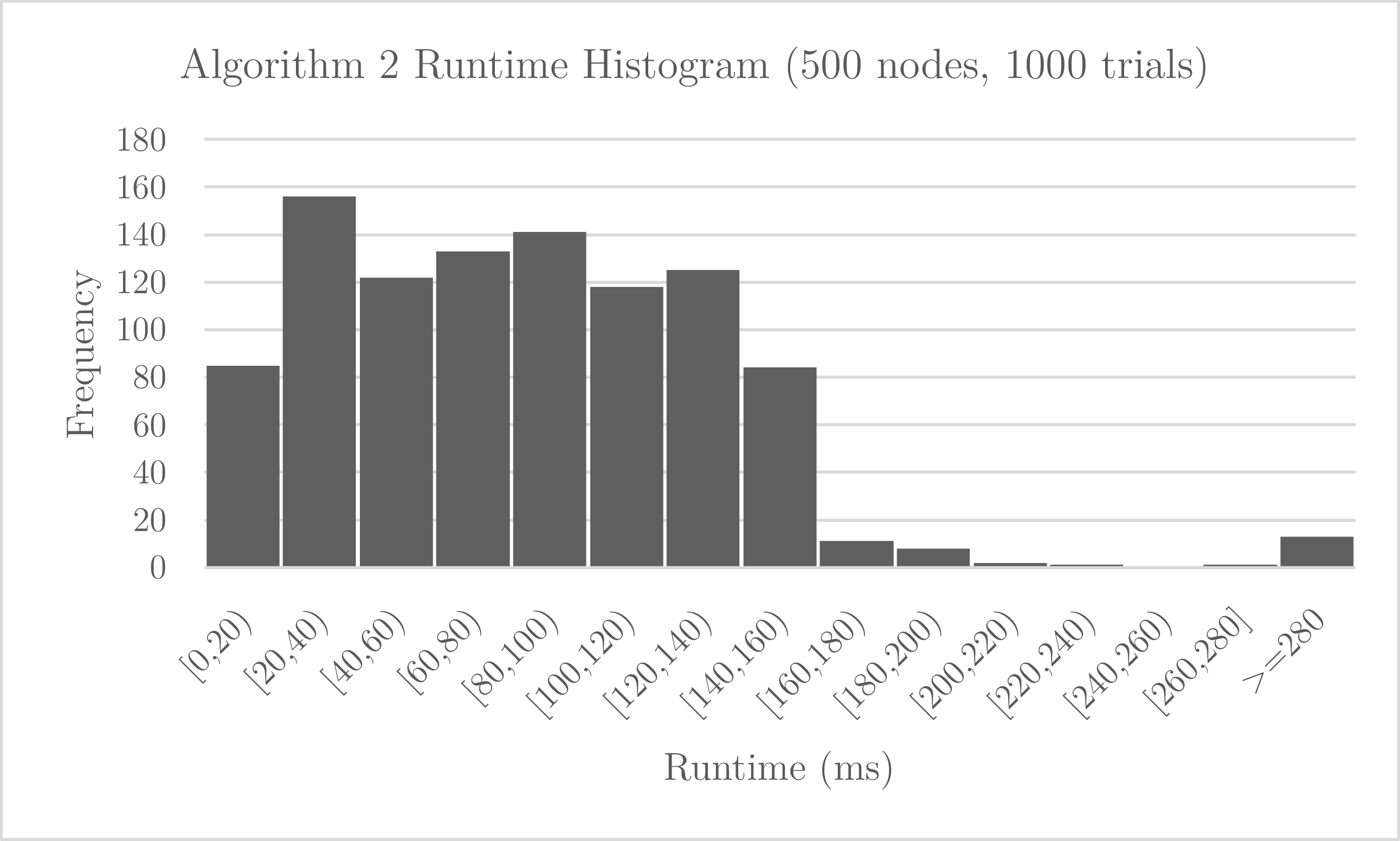}
    \caption{Histogram of the runtime of Algorithm \ref{digraphhash} for $n=500$ nodes over 1000 trials.}
    \label{fig:alg-histogram}
\end{figure}

To determine approximately how often challenging instances occur in a random sample of graphs, we generated 1000 random 500 node graphs and ran Algorithm \ref{digraphhash} on each. Graphs that took longer than 10 seconds were considered challenging. As Figure \ref{fig:alg-histogram} shows, approximately 1.2\% of the test graphs were considered challenging. Although this percentage is fairly low, it is high enough for an adversary to easily generate difficult instances. This may have security implications if the algorithms in this paper are used to hash arbitrary user constructed graphs.

Turning our attention to Algorithm \ref{alg:digraphhashqf} in Figure \ref{fig:alg-all-runtime}, we observe a much steeper curve and more costly running time. This is due to the additional $\log n$ multiplier required for encoding multigraphs as digraphs, as well as the multiple iterations required to reach a fixpoint.

Figure \ref{fig:alg-all-runtime} shows that the runtime of the Merkle digraph hashing algorithm (Algorithm \ref{alg:merklegraphhash}) is slower than simply hashing the entire graph. This means that the nauty library already takes into account the condensation structure when canonizing graphs, so no speedup was observed for computing the canonization of many small graphs vs one large graph. The additional overhead was due to decomposing the large graph into smaller graphs as well as data marshalling and unmarshalling. We therefore conclude that the primary benefit of the Merkle digraph hashing approach is that it allows an incremental construction of hash values.

\section{Use Cases}
\label{usecases}

The Merkle-style algorithm presented here will be useful in cases where recursively hashing cyclic data structures is needed, assuming that the DAG to tree expansion can be tolerated. One benefit of the recursive approach is that hashing can take place at different times, so that data can be built up and hashed over long periods.

An example of a use case for this algorithm is for hashing an interconnected web page system. Tree expansion is tolerable, since it is impossible to distinguish between two equivalently structurally hashed web pages. The recursive property is desirable for performance reasons as web pages build up over longer periods of time. IPFS \cite{ipfs} is an example of a currently available system that uses a Merkle-tree hash for storing web page data, but does not support cycles. This is unfortunate considering that most websites have cycles within their navigation menus.

The algorithm presented in this paper could be used for this use case, modified by the fact that the hyperlinks between web pages occur in a certain order (ie, the edges are labeled), and there may be multiple edges between nodes (ie, a web page may link to another more than once). The labels for the nodes in the graphs will simply be the content of the web pages, with any information about inter-page linkage removed. The inter-page linkage can be restored by examining a node's outgoing edge labels.

Another potential use case is for hashing computer programs as is done in a system like Unison \cite{unison}. In Unison, programs are stored in a database based on their structure. Unison uses the technique given in this paper of hashing entire strongly connected components, but resorts to arbitrarily ordering nodes when the non-recursive hashes between two nodes are equivalent. This presents issues in certain scenarios where nodes have the same non-recursive hashes but do not lie in the same orbit. Systems like Unison could benefit from taking advantage of the canonical labelling algorithm and quotient graph hashing.

Maziarz et al. \cite{Maziarz2021-tq} presented a system for hashing programs modulo alpha equivalence. Looking beyond alpha equivalence, e-graphs are a concise way of representing all the possible ways rewrite rules that can be applied to a program fragment. Willsey et al. \cite{Willsey2021-av} showed that e-graphs can be fully saturated in a space efficient manner by constructing a hypergraph representation. Our work could be used to hash this hypergraph, which is a clear extension beyond hashing programs modulo alpha equivalence. This would allow for hashing of a program modulo the application of rewrite rules.

By using the hashing algorithms presented in this paper we can efficiently use graph data as keys in database systems. This unlocks efficient graph querying that takes into account graph isomorphism. One clear application is with molecular database systems, where chemical information is stored in relation to its chemical structure. Using the algorithms presented in this paper, we can efficiently lookup chemical information given an input molecular structure undirected graph.

\section{Related Work}

The Weisfeiler Lehman graph hash \cite{Shervashidze2011-dt} is a kernel method that iteratively hashes and aggregates neighborhoods of each node in an input graph. Isomorphic graphs are guaranteed to result in equal hashes, and a strong guarantee is given that non-isomorphic graphs will have different hashes. This kernel method takes an iteration parameter as input, which changes the result of the hashes computed for different settings of this parameter. Our algorithm gives similar guarantees as the Weisfeiler Lehman graph hash, with the added benefit that there is no requirement for an iteration parameter.

In a 2014 paper, Arshad et al. \cite{Arshad2014-cc} presented an algorithm for hashing directed graphs, given additional inputs of a fixed source node and depth-first search traversal tree. The depth-first search traversal tree breaks the graph up into different edge types (tree, back, forward, and cross edges) depending on search order. The hashing algorithm then uses the traversal numbers to order the nodes and construct the hash. The authors note that isomorphic graphs may hash to different values due to this input dependence on source node and DFS tree. In contrast, the algorithms presented in this paper do not rely on an input DFS tree or node ordering.

In a separate 2018 paper, Arshad et al. \cite{Arshad2018-ij} presented a graph hashing algorithm where the hash of a node depends on the XORing together the hashes of the labels of all of that node's descendants. The hash of the entire graph is then computed by hashing each node concatenated with the accumulated graph hash. This approach is problematic in two different ways. While the node hash heuristic may be admissible in many cases, it is easy to construct two non-isomorphic graphs containing nodes with equal hashes, and therefore also equal graph hashes. Indeed, consider a $k$-clique and a $k$-cycle graph containing nodes with equivalent labels. These two graphs are not isomorphic, but each node has $k-1$ descendants, each of which are identically labelled. This also shows that the hash of the entire graph are equal. Another problematic aspect of their approach lies with the dependence on DFS order. Two isomorphic graphs may have different graph hashes depending on the DFS order.

\section{Conclusion}

In this paper we have presented multiple algorithms for hashing directed graphs. Previous algorithms given in the area of directed graph hashing require additional input parameters (such as iteration count or order traversal) which our algorithm does not require. As a new cryptography primitive, our algorithm opens doors for applications that require fingerprinting directed graphs and more complex data structures. Our algorithm draws the connection between graph hashing, graph canonization and orbit partitioning. Although our algorithm has exponential run-time in worst case scenarios, many real world graphs have nodes with unique labels. In such cases the run-time of our algorithms are polynomial. Even in cases where node labels are not unique, our algorithm shows acceptable performance for graphs with hundreds to thousands of nodes.

\backmatter

\section*{Declarations}

Python implementation of algorithms presented is available at \url{https://github.com/calebh/dihash}.


% Generated 4/18/2023

\end{document}